\documentclass[twocolumn,showpacs,
               preprintnumbers,amsmath,amssymb]{revtex4}

\usepackage{bm}

\begin{document}

\preprint{UCB-PTH-06/19}
\preprint{LBNL-61552}
\preprint{hep-th/0609107}

\title{
A New Lorentz Violating Nonlocal Field Theory From String-Theory}

\author{Ori J. Ganor}
\email{origa@socrates.berkeley.edu}
\affiliation{%
Department of Physics,
University of California, Berkeley, CA 94720, U.S.A.  \\
and\\
Theoretical Physics Group,
Lawrence Berkeley National Laboratory,
Berkeley, CA 94720, U.S.A.
}%
\date{November 16, 2006}
\begin{abstract}
A four-dimensional field theory 
with a qualitatively new type of nonlocality is constructed
from a setting where Kaluza-Klein particles probe
toroidally compactified string theory with twisted boundary 
conditions.
In this theory
fundamental particles are not pointlike
and occupy a volume proportional to their R-charge.
The theory breaks  Lorentz invariance
but appears to preserve spatial rotations.
At low energies, it is approximately
$N=4$ Super Yang-Mills theory, deformed 
by an operator of dimension seven.
The dispersion relation of massless modes in vacuum is unchanged,
but under certain conditions in this theory,
particles can travel at superluminal velocities.
\end{abstract}



\maketitle

\def\be{\begin{equation}}
\def\ee{\end{equation}}
\def\bear{\begin{eqnarray}}
\def\eear{\end{eqnarray}}
\def\nn{\nonumber}

\newcommand\secref[1]{Section~\ref{#1}}

\def\Tr{{\text{Tr}}} 
\newcommand\rep[1]{{\bf {#1}}} 

\newcommand{\R}{\mathbb{R}}
\newcommand{\Z}{\mathbb{Z}}

\def\Mst{M_{\text{st}}} 
\def\gst{g_{\text{st}}} 
\def\gYM{g_{\text{ym}}} 

\newcommand\SUSY[1]{{$N={#1}$}}  

\def\hM{{\zeta}} 
\def\hJ{{\hat{J}}} 

\def\mj{{j}} 
\def\xm{{m}} 
\def\xe{{e}} 
\def\xk{{k}} 

\def\vB{{\bf B}} 
\def\vE{{\bf E}} 
\def\vP{{\bf P}} 
\def\vn{{\bf\hat{n}}} 

\def\Vol{{V}} 

\def\vBeff{{\bf B}_{\text{eff}}} 
\def\vEeff{{\bf E}_{\text{eff}}} 

\def\a{\alpha}
\def\b{\beta}
\def\g{\gamma}

\def\Op{{\cal O}} 
\def\fJ{{\it {J}}} 

\def\Mp{M_{\text{p}}} 

\def\Hilb{{\cal H}} 
\def\wE{{\widetilde{E}}} 

\section{\label{sec:intro}Introduction}

Violation of Lorentz invariance at high energies is an
interesting theoretical possibility, and it is important
to explore possible extensions of the Standard Model,
and field theories in general, that incorporate it.
Indeed, there exists a large body of work that covers
various aspects of possible Lorentz violating
extensions of field theories
(see \cite{Coleman:1997xq}-\cite{Cheng:2006us} 
and references therein for a small sample of a vast literature).
Many models of this kind start by adding to the Lagrangian
a Lorentz violating term that is an IR-irrelevant
local operator (so that the low-energy behavior will
be unaffected), and then there arises the question of 
whether a consistent UV-completion exists.

Consider, for example,
adding a local Lorentz violating term to
\SUSY{4} Super Yang-Mills (SYM) theory.
By themselves, terms of conformal dimension $\Delta>4$
lead to a theory that is not UV-complete.
Nonetheless, some examples of UV-complete Lorentz violating
deformations of SYM are known.
One example is SYM on a space of noncommutative geometry
\cite{Connes:1997cr}.
There, at low energy the deformation operator
is a 2-form of dimension $\Delta=6$, which breaks the Lorentz
group to $SO(2)\times SO(1,1).$
Another example is dipole-theory \cite{Bergman:2000cw}
where at low energy the deformation operator is 
a spacetime vector of dimension $\Delta=5$,
which breaks the Lorentz group to $SO(2,1).$
In both examples, UV-completeness is maintained because,
in addition to the leading deformation operator,
the Lagrangian has
an infinite number of nonrenormalizable local terms, which
sum up to renormalizable {\it nonlocal} interactions.
Both examples can be realized in String Theory 
\cite{Douglas:1997fm}\cite{Seiberg:1999vs}\cite{Bergman:2001rw}.
A general discussion of nonlocality and its relation to a consistent UV-completion 
of nonrenormalizable interactions appeared recently
in \cite{Adams:2006sv}.

For phenomenological applications, and also for theoretical
exploration, it would be interesting to have new examples of
Lorentz violating theories that  break $SO(3,1)$ to $SO(3)$,
thus preserving spatial rotations. 
In this letter I propose a string theoretic
construction of such a nonlocal, Lorentz violating field theory.
The theory is a deformation of \SUSY{4} SYM, and the deformation
parameter has the dimensions of volume.
This defines a new kind of nonlocality which is fundamentally
different from the two examples mentioned above.

The construction, which involves brane probes in type-II
string theory, is presented in \secref{sec:construction}.
In \secref{sec:BPS} BPS bounds on energies of states
with electric and magnetic fluxes are presented and interpreted.
\secref{sec:disc} concludes with a discussion
of various novel effects in this theory,
including superluminal velocities.

\section{\label{sec:construction}Construction}

The formulation of the new nonlocal field theory
is inspired by Douglas and Hull's construction of
gauge theories on a noncommutative torus \cite{Douglas:1997fm}.
Douglas and Hull started with a compactification of type-IIA
string theory on a small $T^2$ and considered $n$ coincident
D0-branes in the limit where the area of the $T^2$ 
approaches zero.
This setting
is T-dual to type-IIA on a large $T^2$ with $n$ D2-branes,
and can be described by a $U(n)$ Super Yang-Mills (SYM) theory
at low energy. But if an NSNS 2-form flux $B$ is turned on
along the $T^2$, 
T-duality does not map a small $T^2$ to a large one.
Rather, as Douglas and Hull argued, in an appropriate limit
the D2-branes are described by a field theory
with nonlocal interactions. It is a deformation
of SYM theory that can be formally interpreted as a field theory
on a torus whose coordinates are noncommutative.

Let us now turn to our construction.
Start with type-IIA string theory on $T^3$,
and let the compactification radii be $R'_1, R'_2, R'_3$.
Denote the type-IIA string scale by $\Mst'$,
and the type-IIA string coupling-constant by $\gst'$.
Now add a Kaluza-Klein particle with $n$ units of momentum in
the $1^{st}$ direction, and take the limit
\bear
R'_1 &\rightarrow& 0,
\quad
\Mst' R'_1 \rightarrow 0,
\quad
{\Mst'}^3 R'_1 \rightarrow \infty,
\label{eqn:LimitIa}
\\
\gst' &\rightarrow& \text{finite},
\qquad
\Mst' R'_k \rightarrow \text{finite},
\quad (k=2,3).
\label{eqn:LimitId}
\eear
An appropriate U-duality transformation transforms
this setting to a configuration of 
$n$ uncompactified D3-branes. Specifically,
T-duality
in the $1^{st}$ direction, followed by S-duality,
followed by T-duality in the $2^{nd}$ and $3^{rd}$
directions converts the Kaluza-Klein particle
to $n$ D3-branes compactified on $T^3$ 
with compactification radii
\be\label{eqn:Ui}
R_1 = \frac{1}{{\Mst'}^2 R_1'},
\qquad
R_k = \frac{\gst'}{{\Mst'}^3 R_k' R_1'},
\quad
(k=2,3)
\ee
and type-IIB string scale and coupling-constant
\be\label{eqn:Uii}
\Mst = \Mst' \left(\frac{\Mst' R_1'}{\gst'}\right)^{1/2},
\qquad
\gst = \frac{1}{{\Mst'}^2 R_3' R_2'}.
\ee
Note that the original type-IIA coupling-constant is 
given, in terms of the type-IIB parameters
$\gst, R_1, R_2, R_3,$ by
$$
\gst' = \left(\frac{R_2 R_3}{R_1^2\gst}\right)^{1/2}.
$$
In the limit \eqref{eqn:LimitIa}-\eqref{eqn:LimitId},
the type-IIB $T^3$ becomes large,
\bear
\Mst R_1 &=& \frac{1}{(\gst' \Mst' R_1')^{1/2}}\rightarrow\infty,
\\
\Mst R_k &=& 
\frac{1}{\Mst' R_k'}
\left(\frac{\gst'}{\Mst' R_1'}\right)^{1/2}\rightarrow \infty
\quad
(k=2,3),
\nn\\ &&
\eear
while the type-IIB string coupling-constant and the shape of
the type-IIB $T^3$ stay fixed,
$$
\gst\rightarrow\text{finite},
\qquad
\frac{R_k}{R_1} = 
\frac{\gst'}{\Mst' R_k'}\rightarrow\text{finite},
\quad
(k=2,3).
$$
Thus, in the limit \eqref{eqn:LimitIa}-\eqref{eqn:LimitId}, 
the Kaluza-Klein particle is described at low-energy by $U(n)$
\SUSY{4} SYM with finite coupling-constant $\gYM^2 = 2\pi\gst$,
compactified on a large $T^3$ of
radii $R_1, R_2, R_3.$

Similarly to Douglas and Hull's $B$-field flux 
\cite{Douglas:1997fm},
we now add an obstruction that will prevent U-duality from
producing a large $T^3$ on the type-IIB side.
Unlike Douglas and Hull's construction, however,
our obstruction will not be a flux but a 
{\it geometrical twist.}
A geometrical twist in the $1^{st}$ 
direction is defined as follows.
Start with flat space $\R^{9,1}$,
and let $t, x_1,\dots,x_9$
be Minkowski coordinates so that the metric is
$$
ds^2 = -dt^2 + \sum_{i=1}^9 dx_i^2.
$$
Now pick a constant matrix $\hM'\in so(6)$ and
make the global identification
\bear
x_1 &\sim& x_1 + 2\pi R'_1,
\nn\\
x_{a+3} &\sim& \sum_{b=1}^6
{\left\lbrack \exp (2\pi\hM')\right\rbrack^b}_a x_{b+3}
\qquad
(a=1,\dots,6).
\label{eqn:Melvin}
\eear
In other words, we mod out $\R^{9,1}$ by a discrete
group generated by a simultaneous translation in the $1^{st}$
direction and rotation in directions $4,\dots,9$.
This group has no fixed points, 
and can easily be extended to act on the spin-structure
of $\R^{9,1}$, by
taking $\exp (2\pi\hM')$ in a spinor representation of $so(6)$.
(Such spaces have had many theoretical applications
in string theory, a sample of which is listed in
\cite{Melvin:1963qx}-\cite{Alishahiha:2003ru}
and references therein.)

Next, we compactify the $2^{nd}$ and $3^{rd}$ directions
on circles of radii $R'_2, R'_3$ with the usual identifications
$x_k\sim x_k + 2\pi R'_k$ ($k=2,3$) and use the resulting
space as a type-IIA background.

We continue as in the beginning of this section; 
we probe the background with a Kaluza-Klein particle with
$n$ units of momentum in the $1^{st}$ direction, 
and we take the limit \eqref{eqn:LimitIa}-\eqref{eqn:LimitId},
combined with 
\be\label{eqn:LimitM}
\hM\equiv 
\frac{{\gst'}^2}{{\Mst'}^8 {R_1'}^3 R_2' R_3'}\hM' 
= \hM' R_1 R_2 R_3
\rightarrow\text{finite}.
\ee
Here $R_1, R_2, R_3$ are still defined by \eqref{eqn:Ui},
but they are no longer the geometrical compactification radii.
The goal of this paper is to argue that in the limits
\eqref{eqn:LimitIa}-\eqref{eqn:LimitId} and \eqref{eqn:LimitM}
the Kaluza-Klein particle
is described at low-energy (below the string scale)
by a nonlocal field theory
that breaks Lorentz invariance but preserves rotational invariance.
We will see that in the IR limit it can be approximated
by \SUSY{4} SYM deformed by a dimension $\Delta=7$ operator,
the deformation parameter $\hM$ having dimension $(-3)$.
For reasons to be explained in \secref{subsec:Rnonl},
I will refer to this conjectured field theory as
{\it Puffed Field Theory} (PFT).
It will be useful to also consider the case where $R_1, R_2, R_3$
are large but finite:
$R_1, R_2, R_3\gg \Mst^{-1}.$
I will refer to this theory as {\it PFT formulated on $T^3$}
and to $R_1, R_2, R_3$ as the formal 
compactification radii.

\subsection{\label{subsec:susy}Supersymmetry}

Before we proceed to explore the unique properties of PFT,
we have to digress and discuss the conditions on $\hM'$
that are required for PFT to be supersymmetric.
Preserving some amount of supersymmetry is important,
because nonperturbatively, 
the generic background \eqref{eqn:Melvin} can be unstable.
The stability of similar solutions has
been analyzed in \cite{Dowker:1995gb}-\cite{Costa:2001if},
and the generic background can decay 
either by a process of ``bubble-of-nothing'' nucleation
or by a process reminiscent of Schwinger pair-production.
However, these mechanisms do not destabilize the vacuum
if supersymmetry is preserved.

It is not hard to see that
the background defined by the identification
\eqref{eqn:Melvin} preserves $8$ supersymmetry generators if
$\hM\in su(3)\subset so(6)$.
We can choose a coordinate basis where $\hM$ is of the form
\be\label{eqn:Mmat}
\hM = \begin{pmatrix}
0 & \b_1 & 0 & 0 & 0 & 0 \\
-\b_1 & 0 & 0 & 0 & 0 & 0 \\
0 & 0 & 0 & \b_2 & 0 & 0 \\
0 & 0 & -\b_2 & 0 & 0 & 0 \\
0 & 0 & 0 & 0 & 0 & \b_3  \\
0 & 0 & 0 & 0 & -\b_3 & 0 \\
\end{pmatrix}\in so(6).
\ee
Then, $8$ linearly independent supersymmetry generators are
preserved if $\b_1+\b_2+\b_3=0$.
If further $\b_3=0$ then $16$ supersymmetries are preserved.
On the other hand, if for all combinations of $(\pm)$ signs 
$\b_1\pm\b_2\pm\b_3\neq 0$, then no supersymmetry is preserved.
In what follows, unless stated differently, 
I will assume that $\b_1+\b_2+\b_3=0$.
The presence of $n$ units of Kaluza-Klein momentum
in the construction of PFT
breaks additional supersymmetry, and thus
PFT preserves $4$ generators if $\b_1+\b_2+\b_3=0$ 
and $8$ if $\b_1=-\b_2$ and $\b_3=0$.

\subsection{\label{subsec:Rnonl}R-charge and nonlocality}

What does {\it Puffed Field Theory} (PFT) look like?
I do not know the full Lagrangian description
of PFT, but it is possible to make several observations
without a full Lagrangian.
In \secref{sec:BPS} exact results
for some low-lying energy states are presented,
and in \secref{sec:AddProp}
a Lagrangian description up to order $O(\hM)$ is discussed.
These results suggest that PFT 
is a nonlocal theory with a unique structure of nonlocality.
In a nutshell, it can be summarized as follows:
{\it R-charge in PFT carries an intrinsic volume
proportional to $\hM$.}

This means the following.
In pure \SUSY{4} SYM, the R-symmetry is $SU(4)$
and R-charge $\hJ$ is an element of the Lie algebra 
$su(4)\simeq so(6).$ The generic parameter $\hM$ of PFT breaks
$SU(4)$ down to its Cartan subalgebra 
$U(1)^3\subset SU(4)$,
because the Cartan subalgebra is the subgroup that commutes with 
a generic element $\hM\in su(4).$
If $\hM$ is such that \SUSY{2} is preserved
(see \secref{subsec:susy}) then the R-symmetry is broken down to
$U(1)\times U(2)\subset SU(4).$
To cast the ``nutshell'' statement above in a formula,
we associate with R-charge $\hJ$,
which is an element of the appropriate unbroken subalgebra
of the Lie algebra $su(4)$, an intrinsic volume
\be\label{eqn:Vol}
\tilde{V} \equiv \tfrac{1}{2}(2\pi)^3\Tr\{\hJ\hM\},
\ee
where both $\hJ$ and $\hM$ are understood as elements
of the Lie algebra $su(4)\simeq so(6)$, and the trace is taken
in the representation $\rep{6}.$
The volume in \eqref{eqn:Vol} can be positive or negative,
which corresponds to opposite orientations.

In pure \SUSY{4} SYM, the six scalars are in the representation
$\rep{6}$ of $su(4)$ and the fermions fit into the representations
$\rep{4}$ and $\rep{\overline{4}}$.
In PFT, 
if we take $\hM$ of the form \eqref{eqn:Mmat},
then objects with the same 
R-charge as the components of the scalars
acquire, according to \eqref{eqn:Vol},
volume factors $\pm\b_1, \pm\b_2, \pm\b_3.$
Similarly, 
objects with the same R-charge as the components of the fermions
acquire volume factors 
$(\pm\b_1\pm\b_2\pm\b_3)/2$.

The heuristic picture advocated in \eqref{eqn:Vol}
can be understood from the construction
of \secref{sec:construction} as follows.
Consider first
the geometric-twist background \eqref{eqn:Melvin},
and define the change of variables
$$
\rho_a e^{i\theta_a}\equiv x_{2a+2} + i x_{2a+3},
\qquad a=1,2,3.
$$
An arbitrary scalar field in the geometry \eqref{eqn:Melvin}
can be expanded in a Fourier series as follows
\bear
\lefteqn{
\phi(t, x_1,\dots, x_9)
= }\nn\\
&&
\sum_{\mj_1, \mj_2, \mj_3} \sum_{k}
\chi_{\mj_1 \mj_2 \mj_3 k}(t,x_2, x_3, \rho_1, \rho_2, \rho_3)
\nn\\ &&
\qquad
e^{i\sum_{a=1}^3 j_a \theta_a}\exp
\frac{i \bigl(k + \sum_{a=1}^3 \b'_a \mj_a\bigr)x_1}{R'_1},
\label{eqn:phiMel}
\eear
where $\chi_{\mj_1 \mj_2 \mj_3 k}$ are arbitrary functions,
and I have taken $\hM'$ to be of the form 
\be\label{eqn:Mmatpr}
\hM' = \begin{pmatrix}
0 & \b'_1 & 0 & 0 & 0 & 0 \\
-\b'_1 & 0 & 0 & 0 & 0 & 0 \\
0 & 0 & 0 & \b'_2 & 0 & 0 \\
0 & 0 & -\b'_2 & 0 & 0 & 0 \\
0 & 0 & 0 & 0 & 0 & \b'_3  \\
0 & 0 & 0 & 0 & -\b'_3 & 0 \\
\end{pmatrix},
\ee
which matches \eqref{eqn:LimitM} and \eqref{eqn:Mmat} if
\be\label{eqn:bbpr}
\b_a\equiv 
\frac{{\gst'}^2}{{\Mst'}^8 {R_1'}^3 R_2' R_3'}\b_a',
\qquad
a=1,2,3.
\ee
Equation \eqref{eqn:phiMel} is the general expression
that satisfies the periodic boundary conditions set 
in \eqref{eqn:Melvin},
and it
can be interpreted as follows \cite{Alishahiha:2003ru}.
Let 
\be\label{eqn:Ppr}
P'_1 \equiv \frac{k + \sum_{a=1}^3 \b_a \mj_a}{R'_1}
\ee
be the Kaluza-Klein momentum
in the $1^{st}$ direction,
and let
\be\label{eqn:hJ}
\hJ = \begin{pmatrix}
0 & -\mj_1 & 0 & 0 & 0 & 0 \\
\mj_1 & 0 & 0 & 0 & 0 & 0 \\
0 & 0 & 0 & -\mj_2 & 0 & 0 \\
0 & 0 & \mj_2 & 0 & 0 & 0 \\
0 & 0 & 0 & 0 & 0 & -\mj_3  \\
0 & 0 & 0 & 0 & \mj_3 & 0 \\
\end{pmatrix},
\ee
be the angular momentum 
matrix.
The unbroken rotation algebra is 
$so(2)\oplus so(2)\oplus so(2)\subset so(6)$,
and I used the embedding in $so(6)$ to express
$\hJ$ as a $6\times 6$ matrix, which will make the notation
more convenient.
Equation \eqref{eqn:phiMel} implies
a linear relation between 
the fractional part of $P'_1 R'_1$
and the angular momentum $\hJ$,
\be\label{eqn:PprRpr}
P'_1 R'_1 = \tfrac{1}{2}\Tr\{\hM'\hJ\}\pmod\Z.
\ee
Now let's inspect \eqref{eqn:PprRpr} after
the U-duality transformation 
\eqref{eqn:Ui}-\eqref{eqn:Uii} is performed,
and after the limits \eqref{eqn:LimitIa}-\eqref{eqn:LimitId} 
and \eqref{eqn:LimitM} are taken.
$P'_1 R'_1$ becomes the {\it effective} number of D3-branes
\textit{$n_{\mbox{\tiny\rm eff}}$},
and we learn from \eqref{eqn:PprRpr} that it is fractional,
formally! The fractional part is given by
\be\label{eqn:neff}
n_{\mbox{\tiny\rm eff}} = \tfrac{1}{2}\Tr\{\hM'\hJ\}
 = \tfrac{1}{2}(R_1 R_2 R_3)^{-1}\Tr\{\hM\hJ\}
\pmod\Z,
\ee
where I have used \eqref{eqn:Ui}-\eqref{eqn:Uii} and
\eqref{eqn:LimitM} to replace $\hM'$ with $\hM.$
The total volume that this effective fractional
number of D3-branes occupies is
\be\label{eqn:TotV}
(2\pi)^3 R_1 R_2 R_3 n_{\text{eff}}
 = \tfrac{1}{2}(2\pi)^3 \Tr\{\hM\hJ\}.
\ee
Thus, a state with R-charge $\hJ$ in PFT heuristically behaves
as if it has an extra finite chunk of D3-brane of
finite volume $4\pi^3 \Tr\{\hM\hJ\}$, as stated in 
\eqref{eqn:Vol}.
Of course, conventional type-IIB string theory doesn't
have such an ``open'' D3-brane.
We will, however, see below that thinking about PFT in this way
is very convenient.

\section{\label{sec:BPS}Electric and magnetic fluxes}

PFT depends on two parameters ---
the dimensionless coupling-constant $\gYM$,
and the dimension $\Delta=-3$ parameter $\hM$, 
which scales like volume.
{}From here until almost the rest of this section,
the discussion will be restricted
to a value of $\hM$ that preserves $8$ supersymmetries
(see \secref{subsec:susy}).
It is then possible to provide a BPS bound on
the energy of a state in PFT (formulated on $T^3$) with given 
momentum, electric and magnetic flux, and R-charge.
The R-charge is taken in the form of
\eqref{eqn:hJ} with $\mj_1 = -\mj_2 \equiv \mj$ and $\mj_3=0$,
and thus is specified by the single integer $\mj$.

The BPS bound can be easily derived from the central charge
of the supersymmetry algebra in the flat supersymmetric background 
defined by the boundary conditions \eqref{eqn:Melvin}.
Note that because of the presence of $\hM'$ in \eqref{eqn:Melvin},
if we define the Kaluza-Klein charge to be an integer,
the central charge will be augmented by a term proportional to R-charge,
as in the numerator of the righthand side of \eqref{eqn:Ppr}.

Before proceeding to the BPS formula, let me note that
the BPS bound can also be derived by
realizing the setting from
the beginning of \secref{sec:construction}
as a decompactification limit of a certain configuration
of charges in type-IIA String-Theory on $T^6.$
Electric and magnetic flux can then be realized
as fundamental string and D1-brane winding numbers.
We also need to realize the R-symmetry charge $\mj$
and the geometrical twist parameter $\b.$
A Kaluza-Klein monopole can do the job for us.

Take a Kaluza-Klein monopole with one unit of charge
dual to Kaluza-Klein momentum in the $6^{th}$ direction.
For large $R_6$, in the absence of other excitations,
it can be described by the Taub-NUT metric:
\bear
\lefteqn{
ds^2 = -dt^2 + \sum_{i=1}^5 R_i^2 dx_i^2
}\nn\\ &&
+ \left(1+\frac{R_6}{2r}\right)^{-1} 
R_6^2(dx_6 - \frac{1}{2}\sin\theta d\phi)^2
\nn\\ &&
+\left(1+\frac{R_6}{2r}\right)
\left\lbrack
dr^2 + r^2 (d\theta^2+\sin^2\theta\,d\phi^2)\right\rbrack.
\label{eqn:TN}
\eear
This metric has an isometry corresponding to the
Killing vector $\partial/\partial x_6$. The isometry has
fixed points at $r=0$ 
where it acts nontrivially as a rotation of the tangent
space. 
By modifying the periodicity conditions
on the coordinates to
\be\label{eqn:ModP}
x_6\sim x_6 + 2\pi N_1 + 2\pi \b' N_2,
\quad 
x_1\sim x_1 + 2\pi N_2,
\quad
(N_1, N_2\in\Z),
\ee
where $\b'\equiv \b/(R_1 R_2 R_3)$,
we can realize the geometry \eqref{eqn:Melvin} near 
the origin $r=0$ of \eqref{eqn:TN},
in the limit $R_6\rightarrow\infty.$
(A similar setting was also used in \cite{Witten:1997kz}
to construct the dual of $(p,q)$ 5-branes,
and in \cite{Cheung:1998wj} to solve the moduli space of certain 
gauge theories with twisted boundary conditions.)
{}From the point of view of type-IIA String-Theory on
$T^6$, all that \eqref{eqn:ModP} does is change the 
asymptotic metric on $T^6$ at infinity.
Specifically, the $T^2$ in the $1^{st}$ and $6^{th}$ 
directions now has a complex structure given by
$\tau = \b' + i(R_1/R_6).$
We can then borrow BPS bounds \cite{Hull:1994ys}
on the mass of a configuration of charges (a ``black-hole'') 
in toroidally compactified type-II string theory 
to construct BPS bounds on the energy of states in PFT.
In particular, the Kaluza-Klein 
momentum in the $6^{th}$ direction is $2\mj.$

I will now present the result, after the appropriate
limits are taken.
Let $\xk_1,\xk_2,\xk_3\in\Z$ be the integer Kaluza-Klein charges
(the quantized units of momenta),
let $\xe_1, \xe_2, \xe_3\in\Z$ be the number of units of
electric flux,
and let $\xm_1, \xm_2, \xm_3\in\Z$ be the number of units of
magnetic flux,  in directions $1,2,3$, respectively.
Set
\be\label{eqn:DefVol}
\Vol \equiv R_1 R_2 R_3,
\ee
and let $\vn_1, \vn_2, \vn_3$ be unit vectors in directions
$1,2,3$, respectively, and define the spatial momentum 
vector
\be\label{eqn:DefP}
\vP\equiv\sum_{i=1}^3\frac{\xk_i}{R_i}\vn_i,
\ee
and spatial electric and magnetic field vectors
\be\label{eqn:DefEB}
\vE\equiv\sum_{i=1}^3 
\frac{\xe_i R_i}{2\pi\Vol}\vn_i,
\quad
\vB\equiv\sum_{i=1}^3 \frac{\xm_i R_i}{2\pi\Vol}\vn_i.
\ee
Then, the BPS bound on the energy turns out to be
\bear
E &=& 2\frac{\Mst^4}{\gst}\mj\b + 
\frac{2\pi^2\Vol^2}{|n\Vol +2\mj\b|}
\left(\frac{\gYM^2}{2\pi}\vE^2 
+\frac{2\pi}{\gYM^2}\vB^2\right)
\nn\\ &&
+|
\vP -\frac{4\pi^2\Vol^2}{|n\Vol+2\mj\b|}\vE\times\vB
|.
\label{eqn:BPSbound}
\eear
The first term on the RHS of \eqref{eqn:BPSbound}
contains the string scale $\Mst$ and is dominant.
This term is to be expected following the picture
sketched at the beginning of \secref{subsec:Rnonl}:
$\mj$ units of R-charge carry an intrinsic volume of 
$2(2\pi)^3\mj\b$,
which in turn accounts for extra energy.
(Note that $(2\pi)^{-3}\Mst^4/\gst$ is the tension of a D3-brane
\cite{Polchinski:1998rr},
and $2(2\pi)^3\mj\b$ is the effective extra volume.)
This term can be eliminated by a redefinition
of the Hamiltonian $H$ of PFT:
\be\label{eqn:RedefH}
H \rightarrow H - \frac{\Mst^4}{2\gst}\Tr\{\hM\hJ\}.
\ee
Since R-charge $\hJ$ is conserved, the extra term commutes with 
the Hamiltonian and therefore has no effect on the dynamics.
The redefinition \eqref{eqn:RedefH} is equivalent to
a time dependent field redefinition, whereby each field 
is multiplied by a time dependent phase proportional to its R-charge.

The remaining terms in \eqref{eqn:BPSbound} reveal
some of the peculiar features of PFT.
Set $\wE\equiv E - 2(\Mst^4/\gst)\mj\b$,
and let us assume that the BPS bound is attained
for some BPS state.
First note that if we set the electric and magnetic fluxes
to zero in \eqref{eqn:BPSbound}
we get $\wE = |\vP|$, and so the dispersion relation
of massless particles in vacuum is unchanged.
Next, note that with the definition
\be\label{eqn:EBeff}
\vEeff \equiv 
\left|
1 + \frac{2\mj\b}{n\Vol}\right|^{-\frac{1}{2}}\vE,
\qquad
\vBeff \equiv 
\left|
1 + \frac{2\mj\b}{n\Vol}\right|^{-\frac{1}{2}}\vB,
\ee
we can rewrite \eqref{eqn:BPSbound} as
\bear
\wE &=& 
\frac{2\pi^2}{n}\Vol\left(\frac{\gYM^2}{2\pi}\vEeff^2 
+\frac{2\pi}{\gYM^2}\vBeff^2\right)
\nn\\ &&
+|
\vP -\frac{4\pi^2}{n}\vEeff\times\vBeff\Vol
|.
\label{eqn:BPSeff}
\eear
This is the same expression as for undeformed \SUSY{4} SYM,
except with $\vE,\vB$ replaced by $\vEeff,\vBeff$.
The first term in \eqref{eqn:BPSeff} is the energy
stored in the electric and magnetic fluxes,
and the second term is the energy associated
with particles that carry momentum, in excess of the
momentum stored in the electric and magnetic fields.
The novelty in PFT is that the quantization condition
on the effective electric and magnetic fluxes,
as given by the combination of 
\eqref{eqn:DefEB} and \eqref{eqn:EBeff}, depends
on the total R-charge of the system, which is
obviously a nonlocal effect.

The singularity in \eqref{eqn:BPSbound} when $n\Vol +2\mj\b=0$
requires some discussion.
In the following, I will relax the supersymmetry restriction
on $\hM$ and allow $\b_1, \b_2, \b_3$ to be generic.
Let $\Hilb(n,\hJ; \hM, \Vol)$ be the sector 
of the Hilbert space of $U(n)$ PFT with
R-charge specified by $\hJ$, as in \eqref{eqn:hJ},
and compactified on a $T^3$ of volume $(2\pi)^3\Vol.$
According to \eqref{eqn:Vol}, this sector
can be thought of as having an effective net D3-brane
volume of 
\textit{$(2\pi)^3 (n \Vol + \frac{1}{2}\Tr\{\hM\hJ\})$}.
This volume can be either bigger or smaller than
the original sum of the volumes of all the D3-branes,
$(2\pi)^3 n\Vol.$
Let $(\mj_1', \mj_2', \mj_3')$ be a set of integers, and
collect them into an $so(6)$ matrix $\hJ'$ as in \eqref{eqn:hJ}.
Then, the above consideration suggests that we should have
an equivalence of Hilbert spaces:
\be\label{eqn:HEq}
\Hilb(n, \hJ; \hM + \hJ'\Vol, \Vol) \simeq
\Hilb(n+\frac{1}{2}\Tr\{\hJ'\hJ\}, \hJ; \hM, \Vol).
\ee
(Note that \textit{$\frac{1}{2}\Tr\{\hJ'\hJ\}$} is an integer.)
As presented,
the construction in \secref{sec:construction} only depends
on the fractional part of $\b_1,\dots,\b_3.$
But it is actually discontinuous in these parameters,
because as we increase, say, $\b_1$ continuously from 
the value of $0$ to $1$ we end up generating $\mj_1$ additional
Kaluza-Klein particles, 
by a mechanism similar to {\it spectral-flow}.
We can modify the construction of \secref{sec:construction}
and place \textit{$(n+\sum_{a=1}^3\mj_a[\b_a])$}
Kaluza-Klein particles instead of $n$
(where $[x]$ denotes the largest integer not exceeding $x$).
Then, \eqref{eqn:HEq} holds.

Now, let us return to the supersymmetric case.
If $n\Vol+2\mj\b=0$, the effective number of 
D3-branes is zero. 
The Hilbert space should then be trivial, and
there are no states with nonzero electric or magnetic flux.
If $n\Vol+2\mj\b<0$, the effective number of D3-branes
is negative, and we should interpret it as $|n\Vol+2\mj\b|$
anti D3-branes. 
(It may, in fact, also make sense to define
the sector as empty if $n\Vol+2\mj\b<0.$)

Equation \eqref{eqn:HEq} suggests that even the $n=0$ PFT
is meaningful, as long as $\Vol<\infty$ and we restrict
to sectors with R-charge that satisfies $\Tr\{\hM\hJ\}>0.$
In the limit $\Vol\rightarrow\infty$,
we see from \eqref{eqn:BPSbound}
that the energy levels with finite electric or magnetic flux
have energies that scale with the volume like $\Vol^{2/3}$
for $n=0$ (compared to $\Vol^{-1/3}$ for $n>0$).
Thus, in the limit $\Vol\rightarrow\infty$ the $n=0$ theory
does not accommodate electric or magnetic flux.
(It might, in fact, become empty altogether in that limit.)

\section{\label{sec:AddProp}Additional properties of PFT}

This section is devoted to a few additional
observations and conjectures regarding the properties of PFT.
PFT is a deformation of \SUSY{4} SYM,
and the deformation parameter is $\hM.$
By construction, this parameter
is in the adjoint representation
$\rep{15}$ of the R-symmetry group $SU(4)$, 
and it has dimension $\Delta=-3.$

The transformation of $\hM$ under the $SO(3,1)$ Lorentz
group is less clear. The construction in \secref{sec:construction}
singles out both the time direction and the $1^{st}$ direction.
However, both the heuristic picture of \secref{subsec:Rnonl}
as well as the BPS formula \eqref{eqn:BPSbound}
suggest that $\hM$ is the $123$ component of a 3-form.
If this conjecture is true, 
PFT preserves the $SO(3)$ symmetry of spatial rotations, and $\hM$
transforms under $SO(3,1)$ as the time component of a 4-vector.
Another argument for $SO(3)$ symmetry is that the U-duality 
transformation that was used in the construction of PFT
in \secref{sec:construction},
after \eqref{eqn:LimitIa}-\eqref{eqn:LimitId},
can be applied in the presence of the Kaluza-Klein monopole
\eqref{eqn:TN}. After the duality we then get $n$ D3-branes
at the origin of the Taub-NUT space, and the parameter $\hM$
becomes a nonzero asymptotic value for a component of the
Ramond-Ramond 4-form potential at infinity with indices $1236$
[referring to equation \eqref{eqn:TN}].
But in the presence of such a nonzero boundary condition,
the topology of the Taub-NUT metric implies a nonzero 
Ramond-Ramond 5-form field strength at the origin.
In this way we transform the PFT setting into
$n$ D3-branes sitting at a place 
(the origin of the Taub-NUT space)
with strong Ramond-Ramond
5-form field-strength that has a component with
three indices parallel to the
D3-branes (directions $123$) and two indices perpendicular
to the branes (see also \cite{Chakravarty:2000qd}
for related constructions).
Additional arguments in favor of the $SO(3)$
symmetry of PFT will be presented in \cite{toAppear}, where the 
supergravity dual is constructed using techniques similar
to \cite{Hashimoto:1999ut}\cite{Bergman:2001rw}\cite{
Alishahiha:2003ru}\cite{Gimon:2003xk}\cite{Lunin:2005jy}.

Although PFT is generally a nonlocal theory,
order by order in $\hM$
it has to be describable by a local Lagrangian.
This would be a low-energy expansion.
In particular, to first order in $\hM$ the correction to
the \SUSY{4} SYM Lagrangian density has to be of the form
$\Tr\{\hM\Op\}$, where $\Op$ is a local operator of dimension
$\Delta=7$, and in the adjoint representation of $SU(4)$.
Furthermore, if $\hM$ transforms as the time component of
a 4-vector, $\Op$ must also be the time component of a 4-vector.
In addition, if $\hM$ is taken to preserve \SUSY{2}
supersymmetry (see \secref{subsec:susy}) then,
up to total derivatives, $\Tr\{\hM\Op\}$ 
must commute with the unbroken supersymmetry generators.
These arguments suggest that $\Op$ is in a protected
supermultiplet. 
In fact, the list of local operators in
short supersymmetry multiplets of \SUSY{4} SYM
\cite{D'Hoker:2002aw}
contains a unique
natural candidate for $\Op.$
It is a descendant
of a chiral primary operator of dimension $\Delta=4,$
and is obtained by acting on the chiral primary with
supersymmetry generators six times. The explicit exression
is rather long, and will be presented elsewhere \cite{toAppear}.

Now set $n=1$ in \eqref{eqn:BPSbound} and expand
to first order in $\hM$ to obtain
\bear
\wE &=& |\vP'| +
2\pi^2\Vol\left(\frac{\gYM^2}{2\pi}\vE^2 
+\frac{2\pi}{\gYM^2}\vB^2\right)
\nn\\ &&
+ \Tr(\hM\fJ_\mu)T^{0\mu} + O(\hM^2), 
\label{eqn:OhM}
\eear
where
$$
\vP' \equiv \vP -4\pi^2\Vol(\vE\times\vB),
$$
is the excess momentum in addition to the electromagnetic field,
$T^{\mu\nu}$ is the energy momentum tensor 
of the electromagnetic field,
and the R-symmetry 4-current $\fJ^\mu$ is defined
to have components
\be\label{eqn:hJdef}
\fJ^\mu = (\frac{1}{\Vol}\hJ,\quad\frac{\vP'}{|\vP'|\Vol}\hJ).
\ee
Therefore, in this case the correction $\Op$
reduces to $T^{0\mu}\fJ_\mu.$ 
(In general $\Op$ has more terms
which do not contribute to the BPS state in this discussion.)

S-duality of \SUSY{4} SYM is also preserved by PFT.
It is obvious from \eqref{eqn:Uii} that
the duality $\gst\rightarrow 1/\gst$ follows from
T-duality in directions $2,3$ in the type-IIA setting,
and this duality is not affected by the parameter $\hM.$
A $\theta$-angle can also be turned on by adding
an NSNS 2-form flux in directions $2,3$.

Additional properties of PFT
including a proposal for the supergravity dual of the theory
and an investigation of the UV properties of the theory
will be reported elsewhere \cite{toAppear}.

\section{\label{sec:disc}Discussion}

On large scales, FRW cosmology breaks Lorentz invariance 
down to rotational invariance, and it is natural to wonder
whether this Lorentz violation has a counterpart in
high energy phenomena. 
If such a violation exists at an energy scale $\Lambda$,
then it is quite reasonable to expect $\Lambda$ to
be of the order of the (3+1D) Planck scale $\Mp$,
in which case the effects involve quantum gravity.
It is also possible, however, that $\Lambda\ll\Mp.$
For example, $\Lambda$ could be around the GUT scale.
It is then possible that an approximate description
around that scale involves a Lorentz 
violating quantum field theory that preserves spatial rotations.

The {\it Puffed Field Theory} described in this letter
is conjectured to be a UV-complete field theory which
breaks Lorentz invariance but preserves spatial rotations.
The Lorentz violation is parameterized by $\hM\sim\Lambda^3.$
Although it is not a phenomenologically realistic model
because of, among other things, the high amount of supersymmetry,
it is quite possible that more realistic models 
with like features can be similarly constructed.

At low energy, the theory contains
a Lorentz violating term that has a rather
universal structure: it is proportional to 
a contraction of the energy-momentum tensor
and the R-current, $T^{0\mu}\fJ_\mu$.
Such a term can have two interesting effects.
First, suppose we have a soliton
of characteristic size $r$ and mass $M.$
Inside the soliton $T^{00}\sim M/r^3$,
and the term $\hM T^{00}\fJ_0$
translates into an effective potential of the order of 
$V_0 = \pm M/\Lambda^3 r^3$ for a particle of  R-symmetry charge
of the order of $\pm 1.$
If $M$ is big enough,
the potential might have a bound state.
In a nonrelativistic order-of-magnitude
analysis, the condition for a bound state is
$1\lesssim m |V_0| r^2\sim m M/\Lambda^3 r$, 
where $m$ is the mass of the R-charged particle.
In this model, ignoring other interactions,
there will be a bound state for particles
with positive R-charge and mass 
$m\gtrsim \Lambda^3 r/M.$

Second, in a medium with nonzero R-charge density
$\chi\equiv \langle\fJ_0\rangle$, the  term $T^{0\mu}J_\mu$
is dominated by
$\chi T^{00}$, and this modifies the dispersion relation
of massless particles so that the speed of light in
such a material becomes approximately $1 + (\hM\chi/2).$
Thus, superluminal velocities can be achieved.
(This is also the case in noncommutative geometry,
as was nicely demonstrated in 
\cite{Landsteiner:2000bw}\cite{Hashimoto:2000ys}.)
However,
it has to be mentioned that the interaction of PFT with gravity
is not straightforward, and this issue has to be addressed before
extensions to phenomenology can be discussed in a meaningful way.
See \cite{Arkani-Hamed:2004ar} for some of the issues that
can arise when a theory with a spontaneously broken Lorentz
invariance is coupled to gravity.

The construction presented in this letter
is reminiscent of the construction of 
dipole field theories in 
\cite{Bergman:2001rw}\cite{Motl:2001dj}\cite{Alishahiha:2003ru}.
The difference is that in order to construct a dipole theory
we need to probe the background \eqref{eqn:Melvin}
with a D0-brane, and take an appropriate limit of 
$R_1, R_2, R_3\rightarrow 0$, while in the case of PFT
we are probing the background with a Kaluza-Klein particle.
(See \cite{Hashimoto:2002nr}\cite{Cai:2002sv}
for a sample of literature discussing 
other configurations of brane-probes in Melvin universes.)

The dipole-theory has a {\it linear} non-locality --
R-symmetry charged objects expand to segments of length
proportional to their R-charge.
A similar nonlocality structure exists in field theories
on noncommutative spaces. There, objects expand to segments
in direction transverse 
to their momentum \cite{Sheikh-Jabbari:1999vm}.
The nonlocality of PFT, on the other hand, might be described
as {\it hyperplanar} -- objects acquire a volume proportional
to their R-charge and expand in a spacelike hyperplane.
A theory with a linear non-locality can often be constructed
by defining a noncommutative product for fields.
However, it is harder to construct a theory where the nonlocal
objects are higher dimensional.
Examples of this kind include Open-Membrane (OM) theory,
which is a deformation of the
six-dimensional $(2,0)$ theory,
and the Open D-brane theories which are
deformations of little string theories \cite{Gopakumar:2000ep}.
In these theories there are formally open membrane
or open D-brane excitations on M5-branes or NS5-branes,
respectively.
Another example is ``disc-pole-theory,'' which is also
a deformation of the $(2,0)$ theory \cite{Bergman:2001rw}.
All these theories are deformations of
already mysterious higher dimensional theories.
PFT can formally be classified as an open D3-brane theory,
but it is special in that the open brane is of the same
type (D3) as the underlying branes.
PFT might be easier to study, however, because it is
a deformation of a 3+1D Yang-Mills theory.
Generalizations of PFT to other types of branes, 
such as M5-branes and
M2-branes can be constructed along similar lines.

I will end this letter by mentioning a few recently discovered
new research directions that might be of relevance.

First, as mentioned in \secref{sec:AddProp} and 
will be further explained in \cite{toAppear},
PFT is formally related to the behavior of D3-branes
in regions with strong RR 5-form flux.
There might therefore be a connection between
the Hamiltonian discovered in \cite{Sheikh-Jabbari:2004ik},
which describes spherical D3-branes in pp-waves,
and a sector of PFT (perhaps defined on $S^3$ rather than $T^3$).

Another recent development that is possibly related to PFT 
is an intriguing extension
of classical geometry \cite{Ramgoolam:2003cs}
that contains a nonassociative structure, and could potentially be
parameterized by a 3-form.
(I am grateful to Washington Taylor for suggesting this.)
The definition of PFT requires a parameter $\hM$ which
can be thought of as a component of a spacetime vector, or a dual 3-form.
Thinking about $\hM$ as a 3-form is more natural, since it measures spatial volume.
In a possibly related development, 
the effective theories on D-brane probes of  
nongeometric fluxes that are
U-dual to 3-form fluxes \cite{Shelton:2005cf}
were studied in \cite{Ellwood:2006my},
where it was suggested that they involve
a nonassociative structure.
Lastly, another modification of geometry that 
also involves a spatial 3-form
and superluminal velocities was recently 
described in \cite{Nastase:2006na}.
It remains to be explored whether or not PFT is related
to any of the above mentioned extensions of classical geometry.

\begin{acknowledgments}
It is a pleasure to thank Mina Aganagic, Lisa Dyson, Eric Gimon, 
Akikazu Hashimoto, Petr Ho\v{r}ava, 
Sharon Jue, Bom Soo Kim, Guy Moore, Anthony Ndirango,
Yasunori Nomura, Washington Taylor and Jesse Thaler
for helpful discussions.
I also wish to thank the hospitality of the organizers of
the conference ``M-theory in the City,'' which took place
at Queen Mary University of London in November 2006,
and where I have learned about recent developments in
nonassociative structures in geometry.
This work was supported in part by 
the Center of Theoretical Physics at UC Berkeley,
and in part by the Director, 
Office of Science,
Office of High Energy and Nuclear Physics, 
of the U.S. Department of
Energy under Contract DE-AC03-76SF00098, 
and in part by the NSF under grant PHY-0098840.
\end{acknowledgments}

\section*{DISCLAIMER}
This document was prepared as an account of work sponsored by the United States Government. While this document is believed to contain correct information, neither the United States Government nor any agency thereof, nor The Regents of the University of California, nor any of their employees, makes any warranty, express or implied, or assumes any legal responsibility for the accuracy, completeness, or usefulness of any information, apparatus, product, or process disclosed, or represents that its use would not infringe privately owned rights. Reference herein to any specific commercial product, process, or service by its trade name, trademark, manufacturer, or otherwise, does not necessarily constitute or imply its endorsement, recommendation, or favoring by the United States Government or any agency thereof, or The Regents of the University of California. The views and opinions of authors expressed herein do not necessarily state or reflect those of the United States Government or any agency thereof or The Regents of the University of California.


\end{document}